\documentclass[conference]{IEEEtran}
\IEEEoverridecommandlockouts

\usepackage{cite}
\usepackage{amsmath,amssymb,amsfonts}
\usepackage{graphicx}
\usepackage{textcomp}
\usepackage{xcolor}
\usepackage[most]{tcolorbox}
\usepackage{balance}
\usepackage{enumitem}
\usepackage[utf8]{inputenc}
\usepackage{wrapfig}
\usepackage{hyperref}
\usepackage{listings} 
\usepackage{xcolor}
\usepackage{microtype} 
\usepackage{xspace}
\usepackage{algorithm}
\usepackage{algpseudocode}
\usepackage{stmaryrd}
\usepackage{soul}
\usepackage{color, colortbl}
\usepackage{amsfonts} 
\usepackage{booktabs} 
\usepackage{graphicx, subfigure}
\usepackage{caption}
\usepackage{epigraph} 
\usepackage{framed}
\usepackage{here}
\definecolor{coolblack}{rgb}{0.0, 0.18, 0.39}
\usepackage{multirow}

\def\BibTeX{{\rm B\kern-.05em{\sc i\kern-.025em b}\kern-.08em
    T\kern-.1667em\lower.7ex\hbox{E}\kern-.125emX}}
\begin{document}
\title{How Maintainable is Proficient Code? \\ A Case Study of Three PyPI Libraries}

\author{
    \IEEEauthorblockN{Indira Febriyanti\textsuperscript{*}, Youmei Fan\textsuperscript{*}, Kazumasa Shimari\textsuperscript{*}, Kenichi Matsumoto\textsuperscript{*}, Raula Gaikovina Kula\textsuperscript{$\dagger$}}
    \IEEEauthorblockA{ \textsuperscript{*}Nara Institute of Science and Technology, Nara, Japan\\
    \textsuperscript{$\dagger$}Osaka University, Osaka, Japan}
}

\maketitle

\begin{abstract}
Python is very popular because it can be used for a wider audience of developers, data scientists, machine learning experts and so on.
Like other programming languages, there are beginner to advanced levels of writing Python code. 
However, like all software, code constantly needs to be maintained as bugs and the need for new features emerge.
Although the Zen of Python states that ``Simple is better than complex,'' we hypothesize that more elegant and proficient code might be harder for the developer to maintain.
To study this relationship between the understanding of code maintainability and code proficiency, we present an exploratory study into the complexity of Python code on three Python libraries.
Specifically, we investigate the risk level of proficient code inside a file.
As a starting point, we mined and collected the proficiency of code from three PyPI libraries totaling 3,003 files.
We identified several instances of high proficient code that was also high risk, with examples being imple list comprehensions, `enumerate' calls, generator expressions, simple dictionary comprehensions, and the `super' function. 
Our early examples revealed that most code-proficient development presented a low maintainability risk, yet there are some cases where proficient code is also risky to maintenance.
We envision that the study should help developers identify scenarios where and when using proficient code might be detrimental to future code maintenance activities.
\end{abstract}


\section{Introduction}

In recent years, Python has remained one of the top programming languages among a diverse community of developers, scientists, and machine learning practitioners. 
This popularity is evident on platforms like GitHub, which now boasts over 100 million developers \cite{github2023}. 
As Python's popularity grows, maintaining high-quality code becomes essential, leading contributors from different fields to continuously improve it to keep Python useable in various cases.
In the field of data science and academia, contributors conduct the effort of writing the Pythonic way peaks performance at scale \cite{10.1145/3524610.3527879}. 
Others are developing tools to detect the level of Python proficiency required to comprehend and deal with a fragment of Python code\cite{10.1145/3524610.3527878}.
After implementing modification frameworks, other developers are showing that most files contain more basic competency files and that not every contributor contributes competent code \cite{10043258}. 
In development, some developers also propose using artificial intelligence to parse release notes documentation and automatically recommend code updates to become compatible with new versions \cite{10336306}. 
Hence, as some other developers evolve their work, they observed that refactoring operations by developers often improve code readability in open-source software systems\cite{9978201}. 
Many contributions by the supportive Python community have fostered its success and prosperity. 
By utilizing functionality in applications, this community has further amplified the availability of public libraries on the Python Package Index (PyPI) platform. With the demand for maintainability across over 560 thousand Python libraries on this platform, developers can't predict how complex project maintenance will become as they grow. 
When the code reaches a high level of proficiency and complexity to the point where no one understands the impact of changes in one block on another, progress can be slowed. 
However, the maintainability in Python libraries can lead to difficulties if the high-level proficiency code is unsure whether the code is maintainable.
In this study, we focus on examining the complexity matrix score of Python libraries on PyPI to investigate the risk level of proficient code in terms of maintainability. 
Specifically, we analyze 3,003 files from three PyPI libraries: fpdf2, mpmath, and pytorch-geometric, to evaluate their proficiency levels. Based on the key of the research, we ask a single research question, \textit{What is the proficiency of safe and risky complex code?}
From our preliminary result, we identified instances of high code proficiency that risk maintainability, with most maintenance tasks requiring simple list comprehensions, `enumerate' calls, generator expressions, simple dictionary comprehensions, and the `super' function. 
Our analysis revealed that highly proficient code development generally poses a low risk to maintainability. 

\section{A Case Study of Three PyPI Libraries}

We selected three PyPI libraries, due to their potential for demonstrating varied proficiency levels and ranks of cyclomatic complexity scores within their codebases, consisting of 3,003 files: 1) \textbf{fpdf2}, a PDF creation library for Python; 2) \textbf{mpmath}, a Python library for arbitrary-precision floating-point arithmetic; and 3)\textbf{ PyG} (PyTorch Geometric), a library built upon PyTorch to easily write and train Graph Neural Networks (GNNs).

To evaluate the proficiency levels, we adopted Pycefr tool \cite{10.1145/3524610.3527878} from the prior works which will obtain an evaluation of competency level. We modified the framework to output ``Project, Directory, File, Class, Start Line, End Line and Level,'' resulting in a total of 242,591 cases for three libraries. We then simplified this result to focus on proficient code, categorized as \textbf{Advance }(competency level C1, effective proficient code in competency level) and \textbf{Mastery} (competency level C2, mastery proficient code in competency level). This simplification resulted in around 3,068 cases.

We calculated the complexity score using Radon\footnote{\url{https://pypi.org/project/radon/}}, to understand the difficulties of code block development. Radon would analyzes the AST of a Python program to compute cyclomatic complexity (CC) scores, which is based on the number of decisions in a block of code +1. We modified the framework to output ``Project, Directory, File, Rank, Line Start, and Line End,'' resulting in a total of 10,874 cases for three libraries. 
We then simplified this result to focus only on two categories: \textbf{Safe }(CC scores ranked A, representing blocks of code with the lowest complexity, which are simple, easy to understand, and have a low risk of errors) and \textbf{Risky} (CC scores ranked F, representing blocks of code with the highest complexity, which is highly complex, difficult to understand, and has a very high risk of errors). This simplification resulted in around 8,769 cases.
In the final step, the detailed numbers of cases are obtained for each project in step to Table~\ref{tab:overviewproject}, we mapped the lines of code identified as proficient Advance-Mastery to the Safe-Risky categories derived from the complexity analysis. This resulted in 2,836 connection cases for further analysis of the relationship between code proficiency and complexity to understand code maintainability.

\begin{table}
    \centering
    \caption{Overview of the libraries used in the study}
    \label{tab:overviewproject}
    \begin{tabular}{lrrrrr}
        \toprule
        Analyzed project & \# files & \# C1 & \# C2 & \# A & \# F \\
        \midrule
        fpdf2 & 1,260 & 211 & 213 & 1,756 & 5 \\
        mpmath & 266 & 350 & 55 & 1,056 & 39 \\
        pytorch geometric & 1,477 & 949 & 1,290 & 5,889 & 24 \\
        \bottomrule
    \end{tabular}
\end{table}

\begin{figure}[tb]
    \centering
    \includegraphics[width=0.8\linewidth]{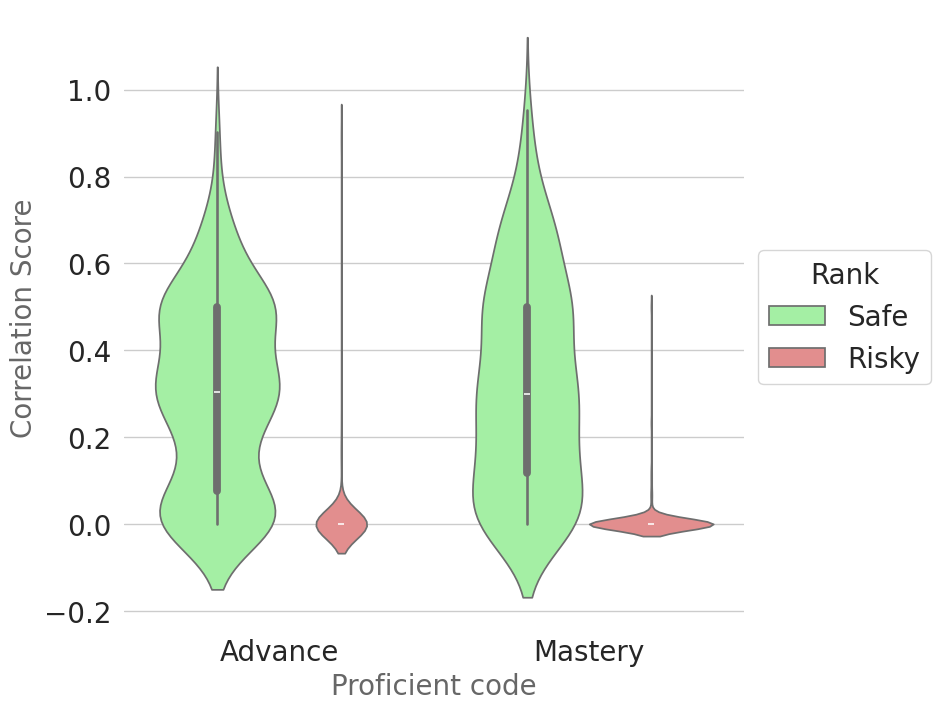}
    \caption{Correlation Scores of three PyPI libraries}
    \label{fig:correlation_score}
\end{figure}

We calculated the correlation score from our findings and visualized it in Figure \ref{fig:correlation_score}. We have four cases with average percentages: 46.61\% as Advance-Safe, 50.25\% as Mastery-Safe, 2.26\% as Advance-Risky, and 0.85\% as Mastery-Risky. This calculation shows that Advance-Safe and Mastery-Safe have the highest percentage, indicating that most of the Proficient code that we found has a low risk of maintainability. This result reveals that proficient code in a large number of cases in the project is mostly maintainable. However, we found a proficient code with a high risk of maintainability of around 3\%, which means even on these three libraries, some parts of the code were hard to maintain. The class code found in those cases is displayed in Table~\ref{tab:topclasses}.

\begin{table}
    \centering
    \scriptsize 
    \caption{Top 5 Code Classes of Proficient and Complexity}
    \label{tab:topclasses}
    \begin{minipage}[t]{0.45\textwidth}
        \raggedleft
        \resizebox{\textwidth}{!}{ 
        \begin{tabular}{lllr}
            \toprule
            Class & Level & Rank & \#Case \\
            \midrule
            Simple List Comprehension & Advance & Risky & 8 \\
            `\texttt{enumerate}' call function & Mastery & Risky & 7 \\
            Generator Expression & Advance & Risky & 2 \\
            `\texttt{zip}' call function & Mastery & Risky & 2 \\
            Super Function & Mastery & Risky & 1 \\
            \midrule
            Super Function & Mastery & Safe & 1,320 \\
            Simple List Comprehension & Advance & Safe & 295 \\
            `\texttt{enumerate}' call function & Mastery & Safe & 168 \\
            `\texttt{zip}' call function & Mastery & Safe & 94 \\
            Generator Function (yield) & Advance & Safe & 75 \\
            \bottomrule
        \end{tabular}}
    \end{minipage}
\end{table}

\section{Conclusion and Future Work}

Our study demonstrates how proficient code maintainability varies between low-risk and high-risk scenarios, providing a clear understanding of high-level code maintainability. 
Interestingly, we observed that some cases with similar classes appeared in both the Risky and the Safe categories. 
This suggests that even similar cases can result in either low or high maintenance requirements. For future work, it would be beneficial to explore the these findings.
We envision that the study should help developers identify scenarios where and when using proficient code might be detrimental to future code maintenance activities.

\section*{Acknowledgments}
This work is supported by JSPS KAKENHI JP20H05706, JP23K16862, JP23K28065, and JST BOOST JPMJBS2423.
\bibliography{bibliography}

\begin{thebibliography}{1}
\providecommand{\url}[1]{#1}
\csname url@samestyle\endcsname
\providecommand{\newblock}{\relax}
\providecommand{\bibinfo}[2]{#2}
\providecommand{\BIBentrySTDinterwordspacing}{\spaceskip=0pt\relax}
\providecommand{\BIBentryALTinterwordstretchfactor}{4}
\providecommand{\BIBentryALTinterwordspacing}{\spaceskip=\fontdimen2\font plus
\BIBentryALTinterwordstretchfactor\fontdimen3\font minus \fontdimen4\font\relax}
\providecommand{\BIBforeignlanguage}[2]{{%
\expandafter\ifx\csname l@#1\endcsname\relax
\typeout{** WARNING: IEEEtranS.bst: No hyphenation pattern has been}%
\typeout{** loaded for the language `#1'. Using the pattern for}%
\typeout{** the default language instead.}%
\else
\language=\csname l@#1\endcsname
\fi
#2}}
\providecommand{\BIBdecl}{\relax}
\BIBdecl

\bibitem{github2023}
\BIBentryALTinterwordspacing
K.~Daigle and GitHub, ``Octoverse: The state of open source and rise of ai in 2023,'' 2023, accessed: 2024-08-05. [Online]. Available: \url{https://github.blog/news-insights/research/the-state-of-open-source-and-ai/}
\BIBentrySTDinterwordspacing

\bibitem{10043258}
\BIBentryALTinterwordspacing
I.~Febriyanti, R.~Kula, R.~Rojpaisarnkit, K.~Kannee, Y.~Nugroho, and K.~Matsumoto, ``Visualizing contributor code competency for pypi libraries: Preliminary results,'' in \emph{2022 29th APSEC}.\hskip 1em plus 0.5em minus 0.4em\relax IEEE Computer Society, dec 2022, pp. 472--476. [Online]. Available: \url{https://doi.ieeecomputersociety.org/10.1109/APSEC57359.2022.00065}
\BIBentrySTDinterwordspacing

\bibitem{10.1145/3524610.3527879}
\BIBentryALTinterwordspacing
P.~Leelaprute, B.~Chinthanet, S.~Wattanakriengkrai, R.~G. Kula, P.~Jaisri, and T.~Ishio, ``Does coding in pythonic zen peak performance? preliminary experiments of nine pythonic idioms at scale,'' in \emph{Proc. of the 30th IEEE/ACM ICPC}, ser. ICPC '22.\hskip 1em plus 0.5em minus 0.4em\relax ACM, 2022, p. 575–579. [Online]. Available: \url{https://doi.org/10.1145/3524610.3527879}
\BIBentrySTDinterwordspacing

\bibitem{10336306}
\BIBentryALTinterwordspacing
N.~Navarro, S.~Alamir, P.~Babkin, and S.~Shah, ``An automated code update tool for python packages,'' in \emph{2023 IEEE ICSME}.\hskip 1em plus 0.5em minus 0.4em\relax IEEE Computer Society, oct 2023, pp. 536--540. [Online]. Available: \url{https://doi.ieeecomputersociety.org/10.1109/ICSME58846.2023.00068}
\BIBentrySTDinterwordspacing

\bibitem{9978201}
\BIBentryALTinterwordspacing
V.~Piantadosi, ``On the evolution of code readability,'' in \emph{2022 IEEE International Conference on Software Maintenance and Evolution (ICSME)}.\hskip 1em plus 0.5em minus 0.4em\relax Los Alamitos, CA, USA: IEEE Computer Society, oct 2022, pp. 597--601. [Online]. Available: \url{https://doi.ieeecomputersociety.org/10.1109/ICSME55016.2022.00082}
\BIBentrySTDinterwordspacing

\bibitem{10.1145/3524610.3527878}
\BIBentryALTinterwordspacing
G.~Robles, R.~G. Kula, C.~Ragkhitwetsagul, T.~Sakulniwat, K.~Matsumoto, and J.~M. Gonzalez-Barahona, ``pycefr: Python competency level through code analysis,'' in \emph{Proc. of the 30th IEEE/ACM ICPC}.\hskip 1em plus 0.5em minus 0.4em\relax ACM, 2022, p. 173–177. [Online]. Available: \url{https://doi.org/10.1145/3524610.3527878}
\BIBentrySTDinterwordspacing

\end{thebibliography}
\bibliographystyle{IEEEtranS.bst}

\end{document}